\pdfoutput=1
\documentclass[a4paper]{article}
\usepackage{amsmath}
\usepackage{graphicx,epstopdf}
\usepackage{cite}
\usepackage{geometry}
\usepackage{array}
\usepackage{amsfonts}
\usepackage{graphicx}
\usepackage[font={small}]{caption}
\usepackage{xcolor}
\usepackage[plain]{fullpage}
\usepackage[colorlinks=true,
            linkcolor=red,
            urlcolor=green,
            citecolor=blue]{hyperref}          
\definecolor{RawSienna}{cmyk}{0,0.72,1,0.45}
\definecolor{dgreen}{rgb}{0.0,0.42,0.13}
\definecolor{darkblue}{rgb}{0.0, 0.0, 0.55}
\definecolor{cornellred}{rgb}{0.7, 0.11, 0.11}
\definecolor{calpolypomonagreen}{rgb}{0.08, 0.5, 0.5}

\def\beq{\begin{equation}}
\def\eeq{\end{equation}}
\def\bea{\begin{eqnarray}}
\def\eea{\end{eqnarray}}
\makeatletter
\begin{document}
\title{\LARGE \bf  Complex Scaling in Neutrino Mass Matrix }
\author{{\bf Rome Samanta$^1$\footnote{ rome.samanta@saha.ac.in}, Probir Roy$^2$\footnote{probirrana@gmail.com}, Ambar Ghosal$^1$\footnote{ambar.ghosal@saha.ac.in},}\\
1. Saha Institute of Nuclear Physics, Kolkata 700064, India\\2. CAPSS, Bose Institute, Kolkata 700091, India} 
\maketitle
\begin{abstract}
Using the residual symmetry approach, we propose a complex extension of the scaling ansatz on $M_\nu$ which allows a nonzero mass for each of the three light neutrinos as well as a nonvanishing $\theta_{13}$. Leptonic Dirac CP violation must be maximal while atmospheric neutrino mixing need not to be exactly maximal. Each of the two Majorana phases, to be probed by the search for $0\nu \beta\beta$ decay, has to be zero or $\pi$ and a normal neutrino mass hierarchy is allowed.

\end{abstract}

If $G_i^T M_\nu G_i = M_\nu$ defines a horizontal symmetry for the complex symmetric $M_\nu$ and $U^T M_\nu U=M_d$, where $M_d$ has only real positive diagonal nondegenerate  elements, then another unitary matrix $V=Ud$ also puts $M_\nu$ into a diagonal form, where $d=\rm diag \hspace{1mm} (d_1,d_2,d_3)$ with $ \rm d_{i(i=1,2,3)}=\pm 1$. Moreover, $U^{\dagger}G_iU=d_i$. Each $d_i$  defines a $Z_2$ symmetry and the corresponding $G_i$ is also a representation of that $Z_2$ symmetry. Among eight possible forms of $d_i$, only two can be shown to be independent, taken as $d_2=\rm diag\hspace{1mm} (-1,1,-1),d_3=\rm diag\hspace{1mm} (-1,-1,1)$. Thus the  two independent representations $G_{2,3}$  describe a residual $Z_2 \times Z_2$ flavor symmetry \cite{1,2} in $M_\nu$. In this way we reinterpret  the Simple Real Scaling ansatz\cite{3} in $M_\nu$ as a $Z_2 \times Z_2$ symmetry. We further make a complex extension of this invariance and obtain the corresponding $M_\nu$. Interesting phenomenological consequences follow. Here we sketch our method and present the basic results leaving many details to a future lengthier publication. Throughout we follow the PDG convention. 

 The Simple Real Scaling ansatz \cite{3}  attributes the following structure to the  neutrino mass matrix 
 
\begin{equation}
M_{\nu}^{SRS}=\begin{pmatrix}
X&-Yk& Y\\-Yk&Zk^2&-Zk\\Y&-Zk&Z
\end{pmatrix} \label{e1}
\end{equation} 
with $X$, $Y$, $Z$ as complex mass dimensional quantities and $k$ as a real positive dimensionless scaling factor. It has one vanishing mass eigenvalue with the corresponding eigenvector $(0,\frac{e^{i\frac{\beta}{2}}}{\sqrt{1+k^2}},\frac{ke^{i\frac{\beta}{2}}}{\sqrt{1+k^2}})^T$. The  mixing matrix is  
 
 \begin{equation}
 U^{SRS}=\begin{pmatrix}
c_{12}&s_{12}e^{i\frac{\alpha}{2}}&0\\
-\frac{ks_{12}}{\sqrt{1+k^2}}&\frac{kc_{12}e^{i\frac{\alpha}{2}}}{\sqrt{1+k^2}}&\frac{e^{i\frac{\beta}{2}}}{\sqrt{1+k^2}}\\
\frac{s_{12}}{\sqrt{1+k^2}}&-\frac{c_{12}e^{i\frac{\alpha}{2}}}{\sqrt{1+k^2}}&\frac{ke^{i\frac{\beta}{2}}}{\sqrt{1+k^2}}
\end{pmatrix} \label{e2}
 \end{equation}
 with an arbitrary $\theta_{12}$ and Majorana phases $\alpha,\beta$. Now $G_{2,3}$  can be calculated from $Ud_{2,3}U^{\dagger}$ to be
\begin{equation}
G_2^k=\begin{pmatrix}
-\cos 2\theta_{12} & \frac{k\sin\theta_{12}}{\sqrt{1+k^2}}& -\frac{\sin\theta_{12}}{\sqrt{1+k^2}}\\
 \frac{k\sin\theta_{12}}{\sqrt{1+k^2}}&\frac{k^2\cos2\theta_{12}-1}{1+k^2}&\frac{-k(\cos2\theta_{12}+1)}{1+k^2}\\
  -\frac{\sin\theta_{12}}{\sqrt{1+k^2}}&\frac{-k(\cos2\theta_{12}+1)}{1+k^2}&\frac{\cos 2\theta_{12}-k^2}{1+k^2}
\end{pmatrix}, G_3^{scaling}=
\begin{pmatrix}
-1&0&0\\0&\frac{1-k^2}{1+k^2}&\frac{2k}{1+k^2}\\
0&\frac{2k}{1+k^2}&\frac{k^2-1}{1+k^2}
\end{pmatrix}.\label{e3}
\end{equation}
The form of $U^{SRS}$ in (\ref{e2}) implies a vanishing $s_{13}$. Since this has been experimentally excluded at $>$ $10\sigma$, the SRS ansatz has to be discarded. However, we shall retain $G_{2}^k$ as well as $G_3^{scaling}$ and propose a complex extension.
Our complex extension postulates 
\begin{equation}
(G_3^{scaling})^T (M_\nu)^{CES} G_3^{scaling}=(M_\nu^{CES})^*. \label{e4}
\end{equation}
The corresponding mass matrix $M_\nu^{CES}$ can be deduced to be
 \begin{equation}
 M_\nu^{CES} = \begin{pmatrix}
 x& -y_1 k +i\frac{y_2}{k}&y_1+iy_2\\
 -y_1 k +i\frac{y_2}{k}&z_1-w_1\frac{k^2-1}{k}-iz_2&w_1-i\frac{k^2-1}{2k}z_2\\
 y_1+iy_2 &w_1-i\frac{k^2-1}{2k}z_2&z_1+i z_2
\end{pmatrix}, \label{e5}
\end{equation}

\vspace{2cm}

\noindent
where $x$,  $y_1$,  $y_2$, $z_1$, $z_2$ and $w$ are real mass dimensional quantities. Eq.(\ref{e4}) implies  $U^{\dagger} G_3U^*=\tilde{d}$ or,
\bea 
G_3U^*=U\tilde{d}.\label{e6}
\eea
 Once again, $\tilde{d}_{lm}=\pm\delta_{lm}$ if the neutrino masses $m_{1,2,3}$ are all nondegenerate. The LHS of (\ref{e6}) can be written out as

\begin{equation}
\begin{pmatrix}
-(U_{e1}^{CES})^*&-(U_{e2}^{CES})^*&-(U_{e3}^{CES})^*\\
\frac{1-k^2}{1+k^2}(U_{\mu 1}^{CES})^*+\frac{2k}{1+k^2}(U_{\tau 1}^{CES})^*&\frac{1-k^2}{1+k^2}(U_{\mu 2}^{CES})^*+\frac{2k}{1+k^2}(U_{\tau 2}^{CES})^*&\frac{1-k^2}{1+k^2}(U_{\mu 3}^{CES})^*+\frac{2k}{1+k^2}(U_{\tau 3}^{CES})^*\\
\frac{2k}{1+k^2}(U_{\mu 1}^{CES})^*-\frac{1-k^2}{1+k^2}(U_{\tau 1}^{CES})^*&\frac{2k}{1+k^2}(U_{\mu 2}^{CES})^*-\frac{1-k^2}{1+k^2}(U_{\tau 2}^{CES})^*&\frac{2k}{1+k^2}(U_{\mu 3}^{CES})^*-\frac{1-k^2}{1+k^2}(U_{\tau 3}^{CES})^*
\end{pmatrix}.
\end{equation}
The reality of $(U_{PMNS})_{e1}$ rules out the possibility $(\tilde{d_i})_{11}=1$. Now there are four cases: $\tilde{d}_a \equiv  \rm diag.\hspace{1mm} (-1,1,1)$, $\tilde{d}_b \equiv \rm diag.\hspace{1mm} (-1,1,-1)$, $\tilde{d}_c \equiv  \rm diag.\hspace{1mm} (-1,-1,1)$, $\tilde{d}_d  \equiv  \rm diag.\hspace{1mm}(-1,-1,-1)$.

These structures of $\tilde{d}$ and the use of (\ref{e6}) lead to the equations given in the following table.
\begin{center}

 \begin{tabular}{|c|c|}
\hline
Elements of $U^{CES}$ & Constraint conditions\\
\hline
$\mu1$ & $2kU_{\mu 1}^{CES}=(1-k^2)U_{\tau 1}^{CES}-(1+k^2)(U_{\tau 1})^*$\\
$\tau1$ & $2k U_{\tau 1}^{CES}=-(1-k^2)U_{\mu 1}^{CES}-(1+k^2)(U_{\mu 1})^*$\\
$\mu 2$ & $2kU_{\mu 2}^{CES}=(1-k^2)U_{\tau 2}^{CES}+\eta(1+k^2)(U_{\tau 2})^*$\\
$\tau 2$ & $2k U_{\tau 2}^{CES}=-(1-k^2)U_{\mu 2}^{CES}+\eta(1+k^2)(U_{\mu 2})^*$\\
$\mu 3$ & $2kU_{\mu 3}^{CES}=(1-k^2)U_{\tau 3}^{CES}+\xi(1+k^2)(U_{\tau 3})^*$\\
$\tau 3$ & $2k U_{\tau 3}^{CES}=-(1-k^2)U_{\mu 3}^{CES}+\eta(1+k^2)(U_{\mu 3})^*$\\
\hline 
\end{tabular}
\end{center}
\vspace{1cm}
 
These equations  lead to the result that (1) for case a, $\alpha=\pi$, $\beta=0$, (2) for case b, $\alpha=\pi$, $\beta=\pi$, (3) for case c, $\alpha=0$, $\beta=0$ and (4) for case d,  $\alpha=0$, $\beta=\pi$. Further, $\cos\delta=0$ where $\delta$ is the Dirac phase in $U_{PMNS}$. In addition, we have the prediction $\tan \theta_{23}=k^{-1}$ which implies that the atmospheric mixing angle need not  be strictly maximal.
We have taken the $3\sigma$ ranges \cite{4} for the quantities $|\Delta m_{31}^2|$, $\Delta m_{21}^2$, $\theta_{12}$, $\theta_{23}$, $\theta_{13}$   for our phenomenological analysis. We also take  the upper bound  0.23 eV on the sum of the light neutrino masses.\\

\noindent
Our conclusions are the following:\\

\noindent
1) Both types of neutrino mass hierarchy are now allowed.\\

\noindent
2) For normal hierarchy, the lightest mass $m_1$ ranges from $10^{-4}$ eV to 0.07 eV and for inverted hierarchy the lightest mass $m_3$ ranges from $10^{-4}$ eV to 0.068 eV.\\

\noindent
3) For both  hierarchies, the quantity $|m_{ee}|$ of relevance to $0\nu\beta\beta$ decay can reach upto the value 0.14 eV which will be probed by GERDA phase II data. 
\section*{Acknowledgements}

The work of RS is  supported by the Department of Atomic Energy (DAE), Government of India. PR acknowledges support as a Senior Scientist from the Indian National Science Academy.

{} 


\begin{thebibliography}{}
\bibitem{1}  C.~S.~Lam,
  Phys.\ Lett.\ B {\bf 656}, 193 (2007).
\bibitem{2}   C.~S.~Lam,
  Phys.\ Rev.\ Lett.\  {\bf 101}, 121602 (2008).
\bibitem{3}  L. Lavoura, Phys. Rev. D62, 093011(2000). W. Grimus and Lavoura, J. Phys. G31, 683(2005). R.N Mohapatra and W. Rodejohann, Phys. lett. B644, 59 (2007)  
\bibitem{4} 
  M.~C.~Gonzalez-Garcia, M.~Maltoni and T.~Schwetz,
  arXiv:1512.06856 [hep-ph].
 \end{thebibliography}
\end{document}